\documentclass[prd,aps,twocolumn,nofootinbib]{revtex4}
\def\lst{LST1}

\def\grb{GRB~221009A}

\def\gr{$\gamma$-ray}
\usepackage{graphicx} 
\usepackage{color}
\usepackage{hyperref}

\begin{document}

\title{Pair echo model of CTAO/LST-1 signal from \grb}
\author{Andrii Neronov$^{1,2}$, Ievgen Vovk$^3$, T\'en\'eman Keita$^4$, Davide Miceli$^{5,6,7}$}

\affiliation{
$^1$Universit\'e Paris Cit\'e, CNRS, Astroparticule et Cosmologie, 75006 Paris, France\\
$^2$Laboratory of Astrophysics, \'Ecole Polytechnique F\'ed\'erale de Lausanne, 1015 Lausanne, Switzerland\\
$^3$ Institute for Cosmic Ray Research, The University of Tokyo, 5-1-5 Kashiwa-no-Ha, Kashiwa City, Chiba, 277-8582, Japan\\ 
$^4$ Universit\'e Paris-Saclay, Universit\'e Paris Cit\'e, CEA, CNRS, AIM, 91191, Gif-sur-Yvette, France\\
$^5$ Gran Sasso Science Institute, Viale Francesco Crispi 7,
67100 L'Aquila, Italy\\
$^6$ Istituto Nazionale di Fisica Nucleare (INFN), Sez. Padova, Via Marzolo 8,35131, Padova, Italy\\
$^7$ INAF - Osservatorio Astronomico di Brera, Via Emilio Bianchi 46, 23807, Merate, Italy
}

\begin{abstract}
Interactions of very-high energy \gr s during their propagation through the intergalactic medium result in a delayed secondary ``pair echo'' signal. Detection of such pair echo can be used for measurements of intergalactic magnetic fields (IGMF) in the voids of the Large Scale Structure of the Universe. We notice that a  combination of data of LHAASO and CTAO/LST-1 telescopes on the \gr\ burst \grb\ is consistent with the pair echo model. The LST-1 measurement of the source signal 1.33 days after the burst can be interpreted as the pair echo produced by very-high-energy $\gamma$-rays propagating through intergalactic medium with magnetic field of the strength $\log(B/\mbox{G})=-16.2\pm 0.7$, if its correlation length $\lambda_B$ is larger than 20~kpc.  If the field correlation length is shorter than 20~kpc, the field strength estimate is modified upwards by a factor $(20\mbox{ kpc}/\lambda_B)^{1/2}$. In the specific case of  short correlation length IGMF that is relic from cosmological phase transitions, the pair echo model of the LST-1 signal implies the field strength in the  $10^{-14}-10^{-12}$~G range. Such magnetic field parameters are consistent with all known constraints on IGMF, including the lower bounds  from \gr\ observations of active galactic nuclei.
\end{abstract}

\maketitle

\section{Introduction}

Magnetic fields in galaxies and galaxy clusters are the result of amplification of a weaker ``seed'' field of uncertain origin through the magnetic flux conserving compression and action of different types of dynamos~\citep{1994RPPh...57..325K, 2001PhR...348..163G, 2002RvMP...74..775W, 2008RPPh...71d6901K}. The required seed field may be produced by various galactic phenomena~\citep{2002RvMP...74..775W,2011ApJ...729...73M,2021MNRAS.502.5726G} or it can be of a primordial origin -- a relic from the first moments of existence of the Universe~\citep[see][for a review]{2013A&ARv..21...62D}. These two possibilities may be distinguished observationally: the proto-galactic seed fields would have small volume filling factor~\citep{2021MNRAS.505.5038A,2021MNRAS.502.5726G}, while the primordial fields are expected to fill the entire Universe. The relic fields may be still present in the voids of the Large Scale Structure (LSS) in their initial form, because no compression or dynamo has ever been active there. 

Evidence for existence of non-zero volume-filling magnetic fields in the voids of the LSS has been previously reported based on the observations of \gr\ telescopes \cite{2010Sci...328...73N}. Very-High-Energy (VHE; $\mathrm{\gtrsim 100~GeV}$) \gr s propagating from distant extragalactic sources deposit electron-positron pairs in the voids in result of interactions with microwave and infrared/optical (extragalactic background light; EBL) photon fields~\citep{gould66, jelley66, stecker92}. These electrons and positrons predominantly loose their energy on inverse Compton scattering of Cosmic Microwave Background (CMB) photons and in this way re-generate \gr s of lower energies~\citep{aharonian_coppi, plaga95, 2009PhRvD..80l3012N}. Deflections of trajectories of electrons and positrons by the Intergalactic Magnetic Field (IGMF) present in the voids lead to the differences in the arrival directions of the primary and secondary \gr s and also to the time delay of the secondary signal~\cite{2009PhRvD..80l3012N,neronov10a,2013A&ARv..21...62D}.

\begin{figure}
    \includegraphics[width=\columnwidth]{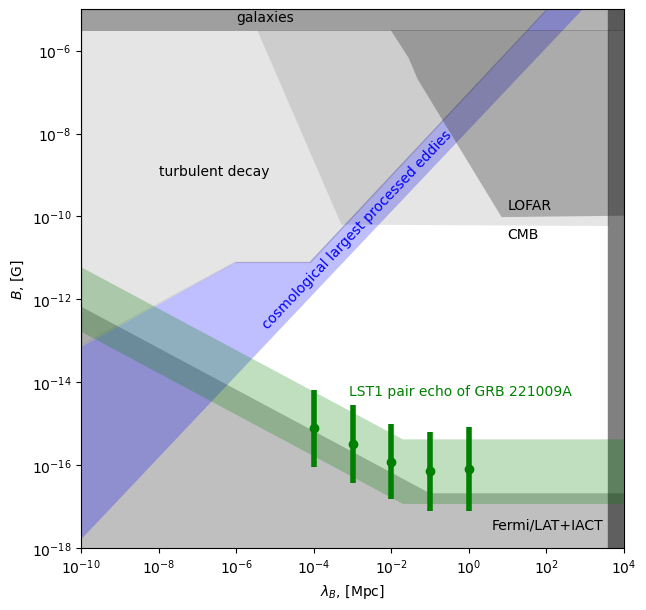}
    \caption{Constraints on IGMF strength $\mathrm{B}$ and coherence length $\mathrm{\lambda_B}$ within the pair echo model of the LST-1 signal. Green shaded data points show 68\% confidence band for $B$ for different assumed correlation lengths. Green band shows the overall dependence of the 68\% band for $B$ on $\lambda_B$. 
    The gray shaded upper bound and labeled stripes / boxes shown known limits on IGMF from \gr\, radio, CMB data as well as from theoretical considerations~\citep[][and references therein]{2023A&A...670A.145A}. Blue color shows the theoretical uncertainty band magnetic field dependent estimate of the largest processed eddy size for IGMF of cosmological origin \cite{Banerjee:2004df,Hosking:2022umv,Brandenburg:2024tyi}. } 
    \label{fig:IGMF}
\end{figure}

Non-detection of such delayed ``pair echo'' signal from \gr\ loud active galactic nuclei (AGN) has been used to derive a conservative lower bound on the void IGMF strength at the level  $B \gtrsim 2\times 10^{-17}$~G~\citep{2011A&A...529A.144T,2023A&A...670A.145A,Blunier:2025ddu} for magnetic field with correlation length larger than $\lambda_B\sim 0.1$~Mpc, see Fig. \ref{fig:IGMF}. Stronger lower bounds have been reported based on non-detection of the IGMF-induced extended emission around AGN emitting in the TeV energy range \cite{2010Sci...328...73N,2018ApJS..237...32A}. These stronger bounds depend on assumptions about the activity period of the TeV source in the AGN. In principle, positive evidence for a pair echo or an extended emission signal can be used to measure the void IGMF parameters, rather than only to constrain the field strength from below. However, no such evidence has been found so far.

An alternative class of sources for pair echo searches are gamma-ray bursts (GRBs), few of which have been detected at TeV photon energies~\citep{2019Natur.575..455M,2019Natur.575..464A,2021Sci...372.1081H,2020GCN.29075....1B}. Contrary to AGNs, the TeV signal from GRBs is transient, possibly providing a more straightforward possibility to disentangle the intrinsic source emission from the pair echo signal.  The pair echos are expected to be present in GRBs that produce \gr s with energies in excess of 0.1~TeV. This is the case for the brightest-of-all-times \grb~\citep{2023ApJ...952L..42L} for which the very-high-energy emission has been detected up to the energies $E_{\gamma_0}>10$~TeV by Large High-Altitude Air Shower Observatory (LHAASO)~\cite{2023Sci...380.1390L}. Several studies comparing \grb\ multi-TeV emission to the lower energy, GeV-band emission detected by Fermi Large Area Telescope (LAT), have suggested a sizable pair echo emission would have been expected in case of IGMF absence. This has been used to derive a lower limit on IGMF strength $\mathrm{B \gtrsim 10^{-19}~G}$~\cite{2024MNRAS.527L..95D,2023ApJ...955L..10H,2024A&A...683A..25V}. Even though this bound is weaker than that derived from the AGN observations, it is an important independent measurement with a different class of sources and hence different possible systematic effects and uncertainties related to the limited knowledge of the primary \gr\ source properties.

In what follows, we use the observations of the \grb\ with LHAASO \cite{LHAASO:2023kyg} and with the first Large Sized Telescope (LST-1) of the Cherenkov Telescope Array Observatory (CTAO) \cite{2019ICRC...36..653C}  to constrain the pair echo signal at higher energies, above 300~GeV. The pair echo signal in this energy range is produced by the highest energy photons with $E_\gamma>10$~TeV \citep{2009PhRvD..80l3012N} and as such probes  somewhat stronger IGMF compared to the Fermi/LAT signal considered in Refs.  \cite{2024MNRAS.527L..95D,2023ApJ...955L..10H,2024A&A...683A..25V}.

\section{The pair echo signal from \grb}

\gr s with energies $E_{\gamma_0}'=E_{\gamma_0}(1+z)> 10$~TeV from \grb\ \cite{LHAASO:2023kyg} at the redshift $z\simeq 0.151$ \cite{2022GCN.32648....1D}  interact on their way through the intergalactic medium producing pairs of electrons and positrons with energies $E_e'\sim E_{\gamma_0}'/2$. Electrons and positrons loose energy on inverse Compton scattering of photons of Cosmic Microwave Background (CMB), producing secondary \gr s with energies \cite{2009PhRvD..80l3012N}
\begin{equation}
 E_\gamma \simeq 0.3\left[\frac{E_{\gamma_0}'}{20\mbox{ TeV}}\right]^2\mbox{ TeV}   
\end{equation}
The cooling distance of electrons with respect to the inverse Compton scattering energy loss is
\begin{equation}
\label{eq:D_e}
D_e\simeq \frac{30}{(1+z)^4}\left[\frac{E_\gamma}{0.3\mbox{ TeV}}\right]^{-1/2}\mbox{ kpc} 
\end{equation}
Trajectories of electrons and positrons are affected by magnetic field present in the intergalactic medium. Deflections of the trajectories of electrons and positrons by IGMF with the strength $B$ and correlation length $\lambda_B$ lead to the displacement of the arrival directions of the secondary \gr s from the source direction and also to the time delay of the secondary signal at the energy $E_\gamma$ by 
\begin{equation}
    \tau_d\simeq \left\{
    \begin{array}{ll}
    \frac{\displaystyle 5\times 10^{6}}{\displaystyle (1+z)^5}
    \left[\frac{\displaystyle E_\gamma}{\displaystyle 0.3\mbox{ TeV}}\right]^{-5/2}&\\ \left[\frac{\displaystyle B}{\displaystyle 10^{-17}\mbox{ G}}\right]^2\mbox{ s}&, \lambda_B>D_{e}\\
    \frac{\displaystyle 5\times 10^{6}}{\displaystyle (1+z)^{2}}
    \left[\frac{\displaystyle E_\gamma}{\displaystyle 0.3\mbox{ TeV}}\right]^{-2}&\\
    \left[\frac{\displaystyle B}{\displaystyle 10^{-17}\mbox{ G}}\right]^2
    \left[\frac{\displaystyle \lambda_B}{\displaystyle  50\mbox{ kpc}}\right]\mbox{ s}&,\lambda_B<D_{e}    
    \end{array}
    \right. 
\end{equation}
The duration of the 10~TeV \gr\ pulse from \grb\ has been $T_{10\ TeV}\sim 10^2$~s \cite{LHAASO:2023kyg}. If the time delay of the echo emission would be $\tau_d\sim T_{10\ TeV}$, the pair echo flux level would be comparable to that of the TeV pulse flux. Otherwise, if $\tau_d\gg T_{10\ TeV}$, the secondary flux is lower by a factor 
\begin{equation}
    \frac{F_{echo}}{F_{10\ TeV}}\sim \frac{T_{10\ TeV}}{\tau_d}
\end{equation}
The favorable range of $E_\gamma, \Delta T$ for detection of the pair echo depends on the parameters of IGMF. The pair echo would not be observable if the followup observation at the energy $E_{\gamma}$ happens after a time delay $\Delta T\gg \tau_d$. Otherwise, the echo may not be detectable if $\tau_d\gg \Delta T\gg T_{10\ TeV}$ because of the too strong suppression of the echo flux.

\section{Modeling of the properties of the pair echo}
\label{sec:crpropa}

The overall fluence of the pair echo signal in the energy range above 0.3~TeV is determined by the primary source fluence in the energy range above 10~TeV. LHAASO WCDA and KM2A detectors have measured the signal from  the \grb\  in this energy range \cite{LHAASO:2023kyg}. The overall time span of the LHAASO signal extends  up to $3000$~s  from the start of the GRB, $T_0$, with the brightest emission  within the time interval  between $T_0+230$~s and $T_0+900$~s. 

LHAASO spectra, corrected for the effect of absorption on EBL \cite{2021MNRAS.507.5144S},  for two time intervals, $T_0+230$~s$<T<T_0+300$~s and  $T_0+300$~s$<T<T_0+900$~s, can be modelled  with a cut-off powerlaw spectral model
\begin{equation}
    \frac{dN}{dE}= N\left(\frac{E}{1\mbox{ TeV}}\right)^{-\Gamma}\exp\left(-\frac{E}{E_{cut}}\right)
\end{equation}
with normalization $N$,  spectral index $\Gamma$ and cut-off energy $E_{cut}$.

For each primary source spectrum model,  we make predictions of the properties of the pair echo signal that depend on the parameters of IGMF. To work out the pair echo model predictions, we use the numerical code \texttt{CRPropa}~v3.2~\citep{2022icrc.confE.978A}. 
\grb\ is assumed to be a point source emitting within a jet with the half-opening angle of $\mathrm{1^\circ}$, as inferred from the LHAASO detection of the "jet break" in the source lightcurve \cite{LHAASO:2023kyg}.  Throughout the simulations, we have injected $\sim 10^4$ primary photons in the $\mathrm{1-100~TeV}$ distributed in energy according to the cut-off powerlaw models and followed propagation of  \gr s and electrons through the intergalactic space.  The intervening IGMF is assumed to be cell-like, i.e. of the constant strength, but randomly orientated from one cell to another. We have explored the the IGMF configurations with field strength in the range $\mathrm{B=10^{-18}-10^{-14}}$~G and cell size $10^{-4}$~Mpc$<\lambda_B<1$~Mpc.  Examples of the resulting pair echo lightcurves  are shown in Fig. \ref{fig:LST-MAGIC} for a range of IGMF parameters and for different energy ranges covered by LST-1.

\section{Constraints on the pair echo from \lst\ telescope}

\begin{figure}
\includegraphics[width=\columnwidth]{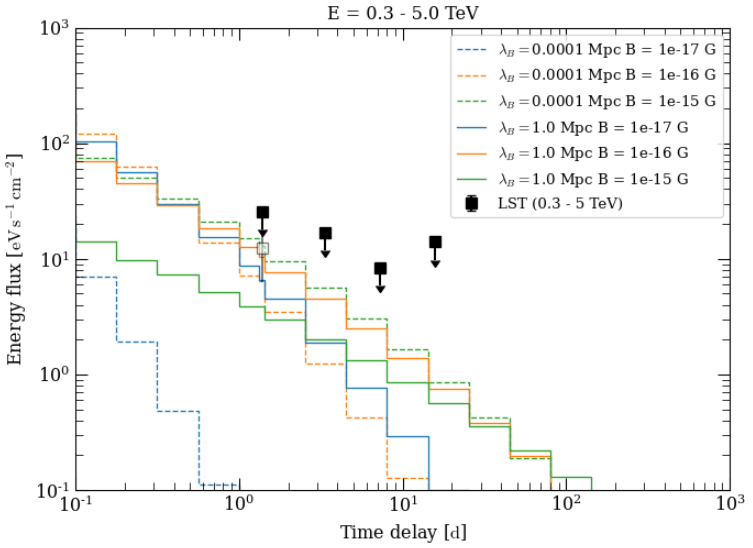}
\includegraphics[width=\columnwidth]{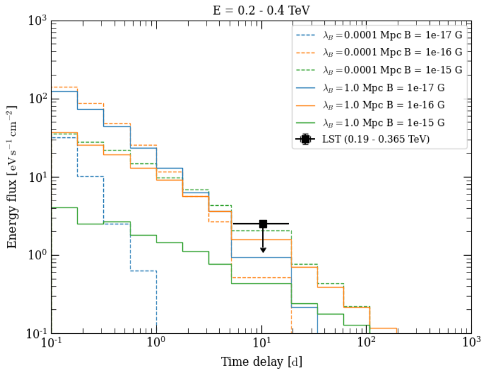}
\caption{Top: \lst\ detection of the source at $T_0+1.33$~day. Bottom: \lst\ upper bound on the source flux for the time interval $T_0+3.33$~d$<T<T_0+17.32$~d. Model lines shod predictions for the pair echo models for different IGMF parameters.}
    \label{fig:LST-MAGIC}
\end{figure}

Strong moonlight has prevented the followup of the \grb\ with Imaging Atmospheric Cherenkov Telescopes (IACT) in the sub-TeV energy range right after the burst occurrence. LST-1 \cite{CTAO-LST:2025wee} has performed observations of the source in the time range from 1.3 to 20 days after $T_0$. Strong moonlight presence during the first observation with LST-1 has resulted in an increase of the energy threshold from the nominal $\mathrm{\sim 20~GeV}$ to $\mathrm{\sim 300~GeV}$. Observations with LST-1 telescope at $T\simeq T_0+1.33$~d have revealed a $4.1\sigma$ excess in the source direction \cite{CTAO-LST:2025wee}. This excess is shown by the empty box data point in the top panel of Fig. \ref{fig:LST-MAGIC}, depicting the LST-1 constraints on the source flux in VHE band. Other observations with LST-1  did not result in significant signal. Only upper limits on the source flux have been derived. These upper limits are shown in Fig. \ref{fig:LST-MAGIC}.

Our numerical modeling of the pair echo signal shows that the  pair echo flux can be at the level of the LST-1 measurement 1.33~d after $T_0$, as is shown in the top panel of Fig. \ref{fig:LST-MAGIC}. It is clear that apart from the pair echo signal, the intrinsic source afterglow can be contributing to the LST-1 detection. However, it is not possible to distinguish the pair echo and the intrinsic afterglow contributions with the limited time coverage and sensitivity of the available VHE band observations.  Given this limitation, we cannot consider the LST-1 detection as an evidence for the pair echo, but we can consider a possibility that the pair echo signal provides an order-of-one contribution to the LST-1 flux measurement 1.33~d after the burst and find parameters of IGMF that are consistent with such a possibility.

\begin{figure}
\includegraphics[width=\columnwidth]{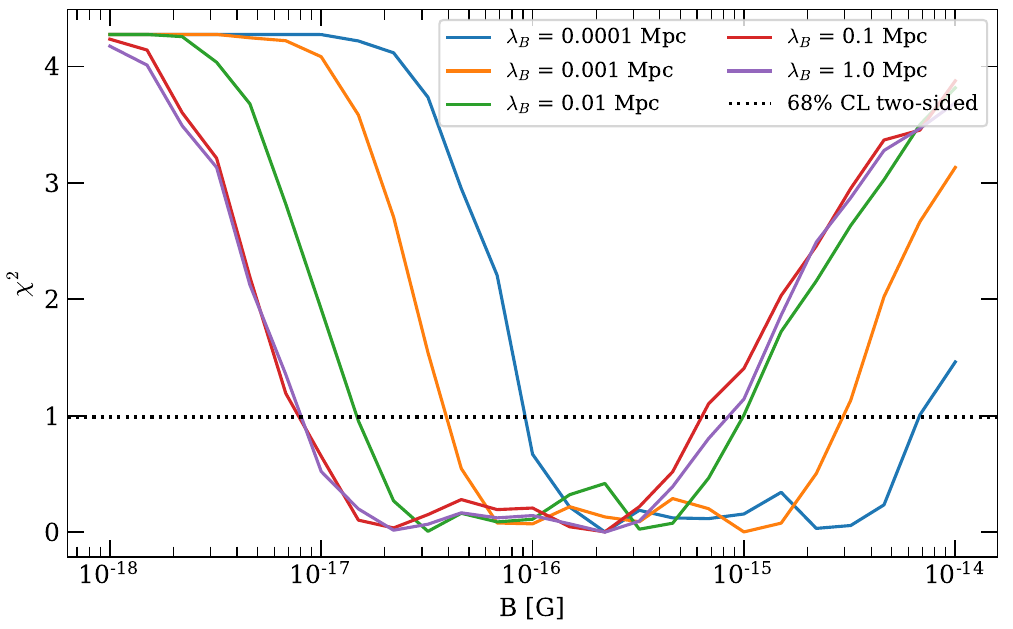}
    \caption{$\chi^2$ profiles of the fits of the joint LHAASO, LST1  data set with the pair echo models for different IGMF strength and coherence scales.}
    \label{fig:Chi2}
\end{figure}

To find the IGMF parameters within the pair echo interpretation of the LST-1 signal, we perform joint fits of the LHAASO and LST-1 data adjusting parameters $E_{cut}, \Gamma, N$ to find the best-fitting model for each magnetic field strength $B$ and correlation length $\lambda_B$.  The spectral data points are fit assuming the gaussian statistics in each energy bin. Fixing  $E_{cut}, \Gamma, N$ to their best-fit values for each combination $B, \lambda_B$, we obtain the profile likelihoods shown in  Fig. \ref{fig:Chi2} as a function of $B$ for different values of $\lambda_B$.  The profile likelihoods have clear minima which means that the LST-1 and LHAASO data provide measurements of the IGMF parameters.  We estimate the 68\% confidence range on $B$ as the range in which the profile likelihood $2\log(L)$ decreases by less than one compared to its maximal value. Dotted horizontal line in Fig. \ref{fig:Chi2} shows the reference level of $\Delta \log(L)$ that determines the boundaries of the errorbars. The estimates of $B$ obtained in this way  for different $\lambda_B$ are shown by the green data points in Fig. \ref{fig:IGMF}. 

These estimates are consistent with the field strength
\begin{equation}
    \log\left[\frac{B}{1\mbox{ G}}\right] =-(16.2\pm 0.7)\left\{
    \begin{array}{ll}
    1,& \lambda_B>20\mbox{ kpc}\\
    \left[\frac{\displaystyle \lambda_B}{\displaystyle 20\mbox{ kpc}}\right]^{-1/2}, &\lambda_B<20\mbox{ kpc.}
    \end{array}
    \right.
\end{equation}
The change of the magnetic field strength estimate at the correlation length $\lambda_B=20$~kpc is explained by the fact that this correlation length is equal to the cooling distance of electrons and positrons producing inverse Compton emission at $E_\gamma\simeq 0.3$~TeV, the energy close to the threshold of the LST-1 observations, see Eq. (\ref{eq:D_e}). The range of allowed IGMF parameters for any $\lambda_B$ is shown by the green band in Fig. \ref{fig:IGMF}.

\section{Discussion}

We have shown that the LST-1 detection of signal from GRB 221009A at $T\simeq T_0+1.33$~d after the burst can be explained by the effect of ``pair echo'' induced by the signal propagation through the intergalactic medium in the presence of intergalactic magnetic field with the strength of the order of $\sim 3\times 10^{-16}$~G (an order-of-magnitude estimate).  This estimate   is consistent with previously derived lower bounds on IGMF stemming from non-detection of the pair echo signal in the  observations of VHE emitting AGN \cite{2023A&A...670A.145A}. The estimated field is weaker than the lower bound reported in Ref. \cite{HESS:2023zwb}, but recent re-analysis of this lower bound \cite{Blunier:2025ddu} shows that the "conservative" lower bound from  this references is relaxed down to the level found in Ref. \cite{2023A&A...670A.145A}. The estimate of the IGMF from GRB 221009A is also consistent  with other known bounds on IGMF from (non)observation of the IGMF induced Faraday rotation of radio signals from AGN \cite{2024arXiv241214825N} and Cosmic Microwave Background bound applicable in the case of IGMF of cosmological origin \cite{2019PhRvL.123b1301J}, shown by different levels of gray shading in  Fig. \ref{fig:IGMF}.  

The IGMF parameters suggested by the pair echo model of the LST-1 signal are consistent with the relic cosmological magnetic field hypothesis. Magnetic fields generated in the early Universe evolve via free turbulence decay following $B\propto \lambda_B^{-\kappa}$ tracks through the $(B,\lambda_B)$ parameter space \cite{Banerjee:2004df,Hosking:2022umv,Brandenburg:2024tyi} with $\kappa=1/2$ in the case of helical magnetic field and $\kappa=5/4$ \cite{Hosking:2022umv} to $\kappa=5/2$ \cite{Banerjee:2004df} in the case of magnetic field with zero helicity. The endpoints of the cosmological evolution lie on the ``largest turbulently processed eddy'' line $B\simeq 10^{-8.5}{\cal C}_M\left[\lambda_B/1\mbox{ Mpc}\right]$~G, where ${\cal C}_{M}\lesssim 1$ is a numerical coefficient that is derived from numerical modeling of the turbulent decay (${\cal C}_M\sim 1$ in the models of Ref. \cite{Banerjee:2004df} while ${\cal C}_M\lesssim 10^{-2}$ was found  in the models of Refs. \cite{Hosking:2022umv,Brandenburg:2024tyi}). The range of uncertainty of the size of the largest processed eddies is shown by the blue shading in Fig. \ref{fig:IGMF}. Intersection of this blue-shaded region with the green band representing the estimate of the IGMF parameters from the GRB221009A data is the locus of possible parameters of the relic magnetic field consistent with the LST-1 signal from GRB221009A. The strength of such field can be between $10^{-14}$~G and $10^{-12}$~G, depending on the correlation length, which is expected to be shorter than parsec. Remarkably, tracing back the evolution of helical relic magnetic field with parameters compatible with the pair echo model of the LST-1 signal, one can find that its parameters at the moment of the Electroweak cross-over in the early Universe may have been those of a field that can produce the observed matter-antimatter asymmetry in the Universe \cite{Boyer:2025jno} through the effect of quantum anomaly coupling the baryon number to the helicity of hyper-magnetic field that existed before the Electroweak transition \cite{Joyce:1997uy,Triangle_anomaly,Giovannini:1997gp,Fujita:2016igl,Kamada:2016_1,Kamada:2016_2}.

It should be noticed that the pair echo model is only one among many possible very-high-energy afterglow of \grb\ \cite{CTAO-LST:2025wee}. A powerlaw extrapolation of the measured TeV-band lightcurve measured by LHAASO~\citep{LHAASO:2023kyg} to the time of the first LST-1 observation gives a prediction for the flux consistent with the LST-1 measurement within the uncertainty range.  Similar conclusion can be drawn from the extrapolation of \grb\ spectrum to the $\mathrm{0.1-1~TeV}$ range based on the lower-energy Fermi/LAT measurements~\citep{2024A&A...683A..25V}.  This shows that a sizeable contribution of the intrinsic source afterglow to the LST-1 signal is possible. The pair echo emission can in principle be disentangled from that intrinsic to GRB based on the characteristic energy dependence of the slope of the signal lightcurve, see Fig. \ref{fig:LST-MAGIC}. However, a study of the energy dependence of the signal in the energy range of interest was not possible for \grb\ because of the strong moonlight that prevented observations with Cherenkov telescopes right after the burst. Nevertheless, such studies will become possible with further GRB detections with  CTAO \cite{2021JCAP...02..048A} and LACT \cite{Zhang:2024ztw}. 

It is clear that even though the pair echo interpretation of the LST-1 signal from GRB221009A is possible, detection of a  signal with day-scale time delay from a  single GRB cannot be considered on its own as a discovery of the pair echo phenomenon. Further detections of similar signal in multiple GRBs and AGN, all consistent with \textit{the same} IGMF parameters would, to the contrary, provide a more convincing evidence for the existence of this phenomenon and a reliable measurement of IGMF parameters. In this sense, our result is demonstrating the potential of the approach to be implemented for next-generation  IACT instruments in construction now, CTAO \cite{2021JCAP...02..048A} and LACT \cite{Zhang:2024ztw}. 

When this paper was ready for publication, we became aware of a publication that reports a similar estimate of IGMF from interpretation of the \lst\ signal of GRB 221009A as a pair echo, \cite{Keita:2026osy}. Contrary to our analysis, the estimate from this paper is only order-of-magnitude in nature (consistent with our results). It does not address the uncertainties of the estimated magnetic field. 

\section*{Acknowledgement}

AN acknowledges the support of  the European
Research Council through the ERC Synergy Grant COSMOMAG
under grant No. 101224803.

\bibliography{refs}
\end{document}